\documentclass[]{aa}
\usepackage{natbib}
\usepackage{graphicx}
\usepackage[varg]{txfonts}
\newcommand{\sect}[1]{Sect.\,\ref{#1}}
\newcommand{\sects}[1]{Sects.\,\ref{#1}}
\newcommand{\fig}[1]{Fig.\,\ref{#1}}
\newcommand{\figs}[1]{Figs.\,\ref{#1}}
\newcommand{\tab}[1]{Table\,\ref{#1}}

\begin{document}

%

\title{Diffuse solar coronal features and their spicular footpoints} 

\titlerunning{Diffuse solar coronal features}
\authorrunning{Milanovi\'{c} et al.}

\author{N.~Milanovi\'{c}, L.~P.~Chitta, H.~Peter}
\institute{Max Planck Institute for Solar System Research, 
           37077 G\"ottingen, Germany, email: milanovic@mps.mpg.de
}

\date{Received ... / Accepted ...}

\abstract
{In addition to a component of the emission that originates from clearly distinguishable coronal loops, the solar corona also exhibits extreme-ultraviolet (EUV) and X-ray ambient emission that is rather diffuse and is often considered undesirable background. Importantly, unlike the generally more structured transition region and chromosphere, the diffuse corona appears to be rather featureless.}
{The magnetic nature of the diffuse corona, and in particular, its footpoints in the lower atmosphere, are not well understood. We study the origin of the diffuse corona above the quiet-Sun network on supergranular scales.}
{We identified regions of diffuse EUV emission in the coronal images from the Atmospheric Imaging Assembly (AIA) on board the Solar Dynamics Observatory (SDO). To investigate their connection to the lower atmosphere, we combined these SDO/AIA data with the transition region spectroscopic data from the Interface Region Imaging Spectrograph (IRIS) and with the underlying surface magnetic field information from the Helioseismic and Magnetic Imager (HMI), also on board SDO.}
{The region of the diffuse emission is of supergranular size and persists for more than five hours, during which it shows no obvious substructure. It is associated with plasma at about 1 MK that is located within and above a magnetic canopy. The canopy is formed by unipolar magnetic footpoints that show highly structured spicule-like emission in the overlying transition region.}
{Our results suggest that the diffuse EUV emission patch forms at the base of long-ranging loops, and it overlies spicular structures in the transition region. Heated material might be supplied to it by means of spicular upflows, conduction-driven upflows from coronal heating events, or perhaps by flows originating from the farther footpoint. Therefore, the question remains open how the diffuse EUV patch might be sustained. Nevertheless, our study indicates that heated plasma trapped by long-ranging magnetic loops might substantially contribute to the featureless ambient coronal emission.}
%
\keywords{Sun: corona, Sun: magnetic fields, Sun: transition region, Sun: UV radiation}

\maketitle

\section{Introduction\label{S:intro}}

\begin{figure*}
\sidecaption
\includegraphics[width=120mm]{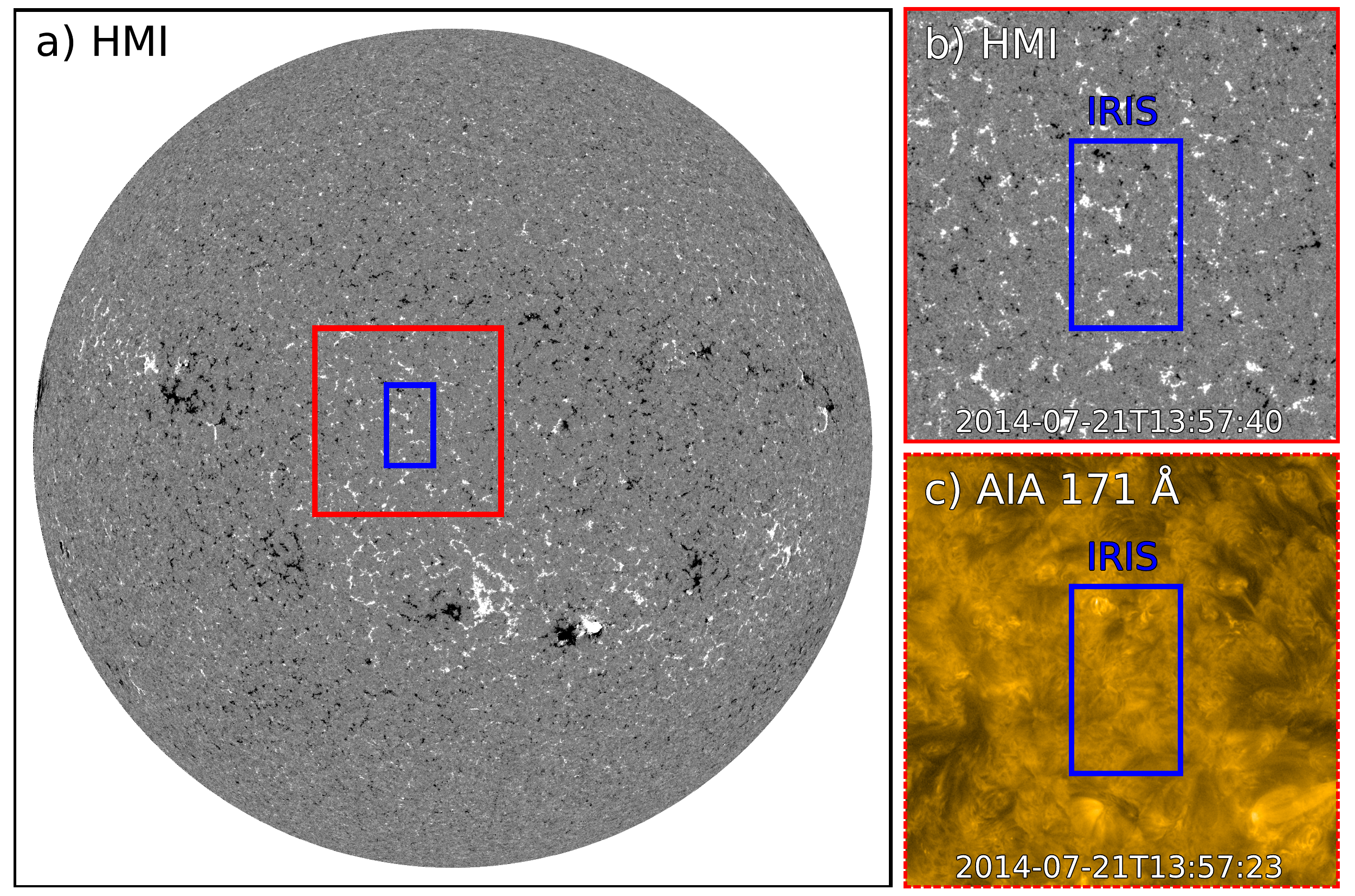}
\caption{Context of our observations. Panel a) shows a full-disk magnetogram from HMI observed around the midpoint of the raster map acquired by IRIS. The IRIS field of view is outlined with a blue box. The field of view indicated by the red box is enlarged in panel b). Panel c) shows the same area as observed with the 171\,\AA\,channel of AIA, plotted in logarithmic scaling. Time stamps that correspond to the beginning of the exposure of HMI and AIA are given with the plots. See \sect{S:obsdata}.}
\label{F:cont}
\end{figure*}

The solar coronal plasma and the associated emission are generally organized into some clearly distinguishable features. This has been recognized in eclipse observations and certainly in the earliest X-ray images taken from space \cite[][]{bray_cram_durrant_loughhead_1991}, followed by images in the extreme-ultraviolet (EUV) light. The brightest features of coronal emission are the active regions, which host multiple coronal loops in which hot plasma is confined by the magnetic field \cite[][]{reale2014}. The large areas that are devoid of emission are called coronal holes \cite[][]{timothy1975}, while the moderately bright regions outside the active regions and coronal holes are referred to as the quiet Sun.

The prominent coronal loops contribute only little to the total coronal emission, even though they are the brightest discrete features, as discussed, for example, by \cite{viall2011}. These authors reported that the radiance of individual loops is often only about 10\%--20\% of the total radiance. In contrast, the diffuse emission that appears to be hazy and without substructure contributes strongly. Early X-ray observations at moderate spatial resolution show this diffuse emission in active and quiet regions \cite[][]{sturrock1996}. This is visible in observations from different instruments and does not depend on the exact temperature of the source region in the corona. When coronal loops are studied, the diffuse emission is often treated as background that needs to be carefully subtracted before the properties of the loops themselves are analyzed \cite[][]{delzanna2003}. Although this background is usually not the main target of the study because it constitutes the bulk of the emission, it is essential to understand how this part of the coronal plasma is heated.

In regions with prominent loops, the diffuse emission might be thought to originate from multiple low-intensity loops that overlap along the line of sight and therefore are basically indistinguishable \cite[][]{williams2020}. However, this would not explain the diffuse emission observed in more quiet areas, as in the study by \cite{sturrock1996}. Regardless of how diffuse the coronal emission might be, however, the chromosphere and transition region below it are highly structured. Chromospheric spicules are one of the prominent features there \cite[][]{beckers1968, depontieu2007} and can also reach transition region temperatures \cite[][]{pereira2014}. It is not understood how this highly structured foundation, starting from the magnetic network at the solar surface, can produce this featureless emission in the corona.
We aim to address this question here. We analyze an observation of the diffuse emission on supergranular scales originating from plasma at close to 1 MK. We combine the coronal observations with observations of the transition region and the underlying surface magnetic field to discuss the formation of this diffuse emission.

\section{Observations and data processing\label{S:obs}}

\subsection{Observational data\label{S:obsdata}}

We used data from three different instruments to cover the atmosphere from the photosphere to the corona. To do this, we used photospheric magnetic field data, UV spectroscopic data for the chromosphere and transition region, and EUV imaging of the corona.

We used a very large, 400-step dense raster acquired by the Interface Region Imaging Spectrograph \cite[IRIS;][]{IRIS} on 21 July 2014 between 11:50 and 15:21 UT (OBSID 3824263396).\footnote{These data are available at https://iris.lmsal.com/.} The target of the observation was a quiet-Sun region near the solar disk center (\fig{F:cont}). With a step size of $0.35\arcsec$ , the raster covers roughly $106\arcsec$ in the east-west direction, and  $171\arcsec$ along the slit (image scale along the slit is $0.17\arcsec$ per pixel). The exposure time per raster step is 30\,s, which ensures a good signal-to-noise ratio. We concentrate here on the transition region emission lines of Si\,{\sc iv} (1394\,\AA) and O\,{\sc iv} (1401\,\AA) and on the Lyman continuum of Si\,{\sc i} around 1395\,\AA\,(see \tab{T.lines}). The spectral scale is 25.6 m\AA\,per pixel, corresponding to about 5.5\,km\,s$^{-1}$.

For a view of the coronal plasma, we used a time series of the full-disk filtergrams at 171\,\AA\,acquired by the Atmospheric Imaging Assembly \cite[AIA;][]{AIA} on board the Solar Dynamics Observatory \cite[SDO;][see \tab{T.lines}]{SDO}. The AIA time series we used starts one hour before the beginning of the IRIS raster and ends one hour after the end of the raster. The plate scale is about $0.6\arcsec$ per pixel, and the cadence is 12\,s. Additionally, we used similar full-disk filtergrams at 131\,\AA, 193\,\AA, 211\,\AA,\,and 304\,\AA to provide context over a range of temperatures (see \tab{T.lines}). These filtergrams were observed around 13:57 UT on the same day, when the IRIS slit crossed the middle of the region of interest.

To study the photospheric magnetic field in the same area on the Sun, we used a time series of the full-disk line-of-sight magnetograms acquired by the Helioseismic and Magnetic Imager \cite[HMI;][]{HMI1, HMI2} on board the SDO.\footnote{All data from AIA and HMI are available at http://jsoc.stanford.edu/.} The HMI data span the duration of the AIA 171\,\AA\,data we considered. The plate scale is about $0.5\arcsec$ per pixel, and the cadence is 45\,s. For additional insights into the magnetic field configuration on larger scales, we also used a daily updated synchronic frame from HMI for 21 July 2014. This data product provides the synchronic map of the magnetic field over the full solar surface, where values of the magnetic field near the poles are filled in \cite[for a full description of the data product, see][]{sun2018}.

\begin{table}
    \centering
    \caption{Lines and EUV bands\label{T.lines}}
   \begin{tabular}{l l c l}
         \hline\hline
         & Line\tablefootmark{a} / band\tablefootmark{b} &  $\log\,T$~[K]\tablefootmark{d} &
         $T$~[MK]\tablefootmark{d} \\
         \hline
         & Si\,{\sc i} continuum   
             & \multicolumn{2}{c}{(low chromosphere)}\\
         IRIS & Si\,{\sc iv} 1393.76\,\AA   &  4.90  &  0.08\\
              & O\,{\sc iv} 1401.163\,\AA   &  5.18  &  0.15\\
         \hline
             & 304\,\AA ~~(He\,{\sc ii})  &  4.92  &  0.08\\
             & 131\,\AA ~~(Fe\,{\sc viii})\tablefootmark{c}  &  5.76  &  0.58\\
         AIA & 171\,\AA ~~(Fe\,{\sc ix})  &  5.93  &  0.85\\
             & 193\,\AA ~~(Fe\,{\sc xii}) &  6.18  &  1.5\\
             & 211\,\AA ~~(Fe\,{\sc xv})  &  6.27  &  1.9\\
         \hline
    \end{tabular}
    \tablefoot{%
    \tablefoottext{a}{Rest wavelengths for Si\,{\sc iv} and O\,{\sc iv} taken from \cite{sandlin1986}.}
    \tablefoottext{b}{For the AIA bands, the primary contributing ions are listed.}
    \tablefoottext{c}{The 131\,\AA\,band also contributes significantly from Fe\,{\sc xxi} around 10\,MK \citep{AIA}, but this is not relevant for our quiet-Sun observations.}
    \tablefoottext{d}{For Si\,{\sc iv} and O\,{\sc iv}, $T$ is the line formation temperature in the MHD model as given by \cite{peter2006}, their Table 1. For the AIA channels, the peak temperature $T$ of the contribution function is listed based on \cite{peter2022}, their Fig.\,9.}}
\end{table}

\subsection{Data processing\label{S:dataproc}}

\begin{figure*}
\centering
\includegraphics[width=\textwidth]{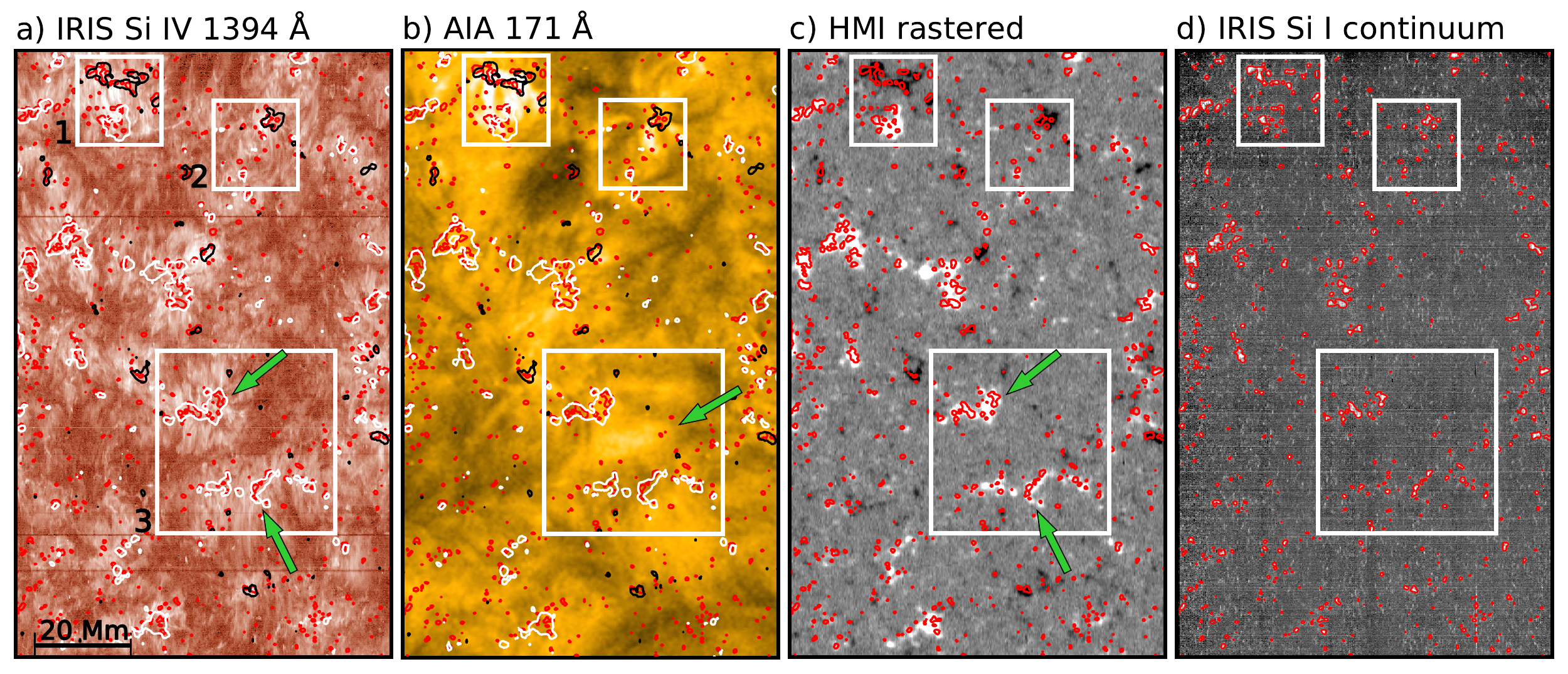}
\caption{Overview of IRIS raster field of view. The individual panels represent a) the total intensity of the Si\,{\sc iv} 1394\,\AA\,emission line obtained from IRIS, b) one image from  AIA in the 171\,\AA\,channel observed near the midpoint of the IRIS raster, c) the rastered line-of-sight magnetogram constructed using data from HMI, which is nearly cotemporal with the IRIS raster (see \sect{S:dataproc}), and d) the Si\,{\sc i} Lyman continuum obtained from IRIS around 1395\,\AA. Panels a) and b) are plotted in logarithmic scale. The black and white contours in panels a) and  b) outline the magnetic field from panel c) at values of $-30$ G and $+30$ G, respectively. The red contours in all panels outline the brightenings in the continuum intensity from panel d). The three white squares and green arrows in all panels highlight different features, as discussed in \sects{S:dataproc} and \ref{S:pattern}.}
\label{F:data}
\end{figure*}

We investigated the properties of the Si\,{\sc iv} (1394\,\AA) line profiles and of the Si\,{\sc i} Lyman continuum. For the Si\,{\sc iv} line, we fit the observed profiles with a single Gaussian by using the curve\_fit function from the Python package SciPy \citep{scipy}. The total line intensity calculated from the parameters of the fit is shown in \fig{F:data}a. During this observation, IRIS did not track the solar rotation, which caused features on the solar surface to appear stretched in the raster. Because the field of view is close to disk center, we assumed that the rotation is fully along the  east-west direction (horizontal in the images). To correct for this during plotting, we adjusted the aspect ratio of all the images that show data from the IRIS raster.

Using the spectra from IRIS, we also obtained a raster map of the Si\,{\sc i} Lyman continuum around 1395\,\AA\,that forms just above the temperature minimum in the low chromosphere \cite[e.g.,][]{vernazza1981}. We show this map in \fig{F:data}d. This image was produced by summing the spectrum over a range between 150~km\,s$^{-1}$ and 420~km\,s$^{-1}$ on the red side of the Si\,{\sc iv} line, covering 49 spectral pixels. We used this approach instead of a Gaussian fitting to reduce the noise of this rather faint continuum level. The continuum image was further processed by flattening visible global trends, and an odd-even pixel pattern in the slit direction, all of which are observed only when working with very low count rates, as we found here in continuum. This process allowed us to detect small brightenings in the chromospheric network in the Si\,{\sc i} continuum. We outline these bright locations with red contours in all panels in \fig{F:data}. A good fraction of these brightenings is expected to be the chromospheric footpoints of the structures observed in the transition region and corona. They are therefore thought to correlate very well with the patches of the enhanced magnetic field in the network.

We prepared the coronal data from AIA by updating the pointing, rescaling the plate scale to exactly $0.6$\arcsec\,per pixel, removing the roll angle, and correcting the observed intensity for the instrument degradation. This was done by using the routines from the Python package aiapy \citep{aiapy}, which is similar to the aia{\_}prep procedure from SolarSoft. To compare IRIS data with the AIA images, we first coaligned the time series of the coronal imagery with the spectral data. With the help of the information given in the data headers, we first projected the AIA data onto the IRIS field of view by using the \mbox{reproject{\_}interp} function from the Python package reproject \cite[part of The Astropy Project;][]{astropy1, astropy2}. We then corrected the AIA images for the effect of solar rotation to track the region of interest targeted by the IRIS observation. We repeated this for each AIA snapshot separately. As for the IRIS data, we assumed that close to disk center, this effect is along the east-west direction, and we shifted the image along the horizontal axis to correct for the rotation effects. The shift was always equal to an integer number of IRIS pixels (which are only a fraction of an AIA pixel). One snapshot in the AIA 171\,\AA\,channel processed in this way is shown in \fig{F:data}b.

Similarly, we projected the magnetic field data from HMI onto the IRIS field of view and corrected them for solar rotation. From the resulting HMI time sequence, we then constructed a rastered line-of-sight magnetogram, as shown in Fig.\,\ref{F:data}c. To do this, we identified the magnetogram closest in time to each IRIS slit position. From this magnetogram, we cut the (vertical) stripe that corresponds to the slit position and added it to the (pseudo) raster. In this way, we built a magnetogram that is nearly cotemporal with the IRIS raster. Because the magnetogram changes during the roughly 3.5 hours while IRIS scans the field of view with its slit, this process is essential to verify the proper alignment between IRIS, HMI, and AIA. Fig.\,\ref{F:data}c shows the good alignment of the red contours that represent the brightening locations in the low chromosphere from IRIS, and the magnetic field concentrations observed in the photosphere by HMI. This match confirms that the different data sets from IRIS and SDO are aligned well.

\begin{figure*}
\sidecaption
\includegraphics[width=120mm]{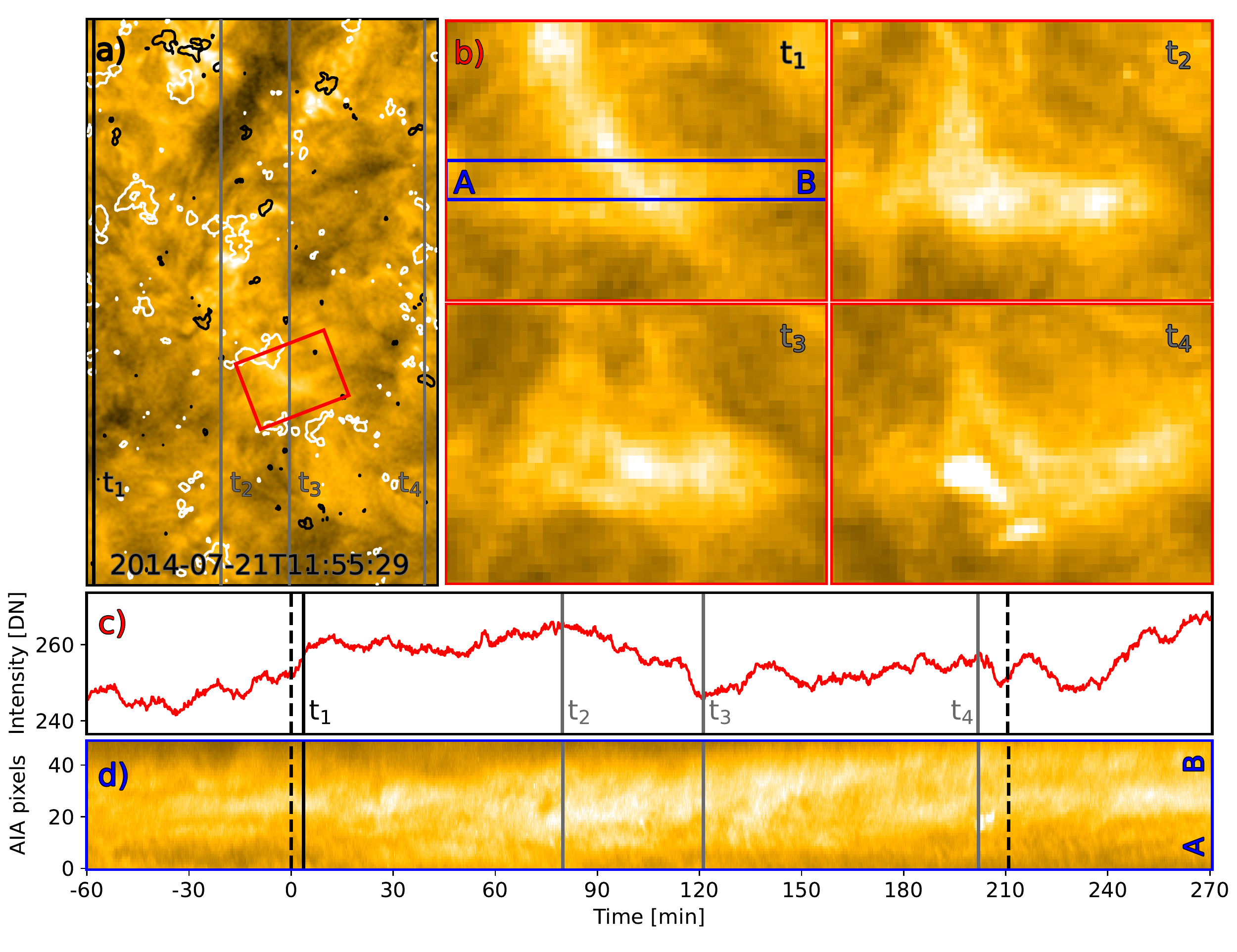}
\caption{Evolution of the diffuse coronal region. Panel a) shows a repetition of \fig{F:data}b displaying the AIA 171\,\AA\,image at time $t_1$ , with black and white contours indicating opposite magnetic patches in the photosphere. The red rectangle highlights the location of the diffuse coronal region. Panel b) shows four snapshots in this rectangle at times $t_1$ to $t_4$ , as indicated in panels c) and d) (plotted in linear scaling). Panel c) shows the variability of the emission averaged within the red rectangle, which does not exceed about\,10\% over more than five hours. Panel d) displays the emission along the slit from A to B, as indicated in panel b). The intensity here is integrated (vertically) between the blue lines. The time axis in panels c) and d) is set to zero when the IRIS raster starts. The IRIS raster ends at about 210 min. The temporal evolution is available in an online movie. See \sect{S:evolution}.}
\label{F:evol}
\end{figure*}

\begin{figure*}
\centering
\includegraphics[width=\textwidth]{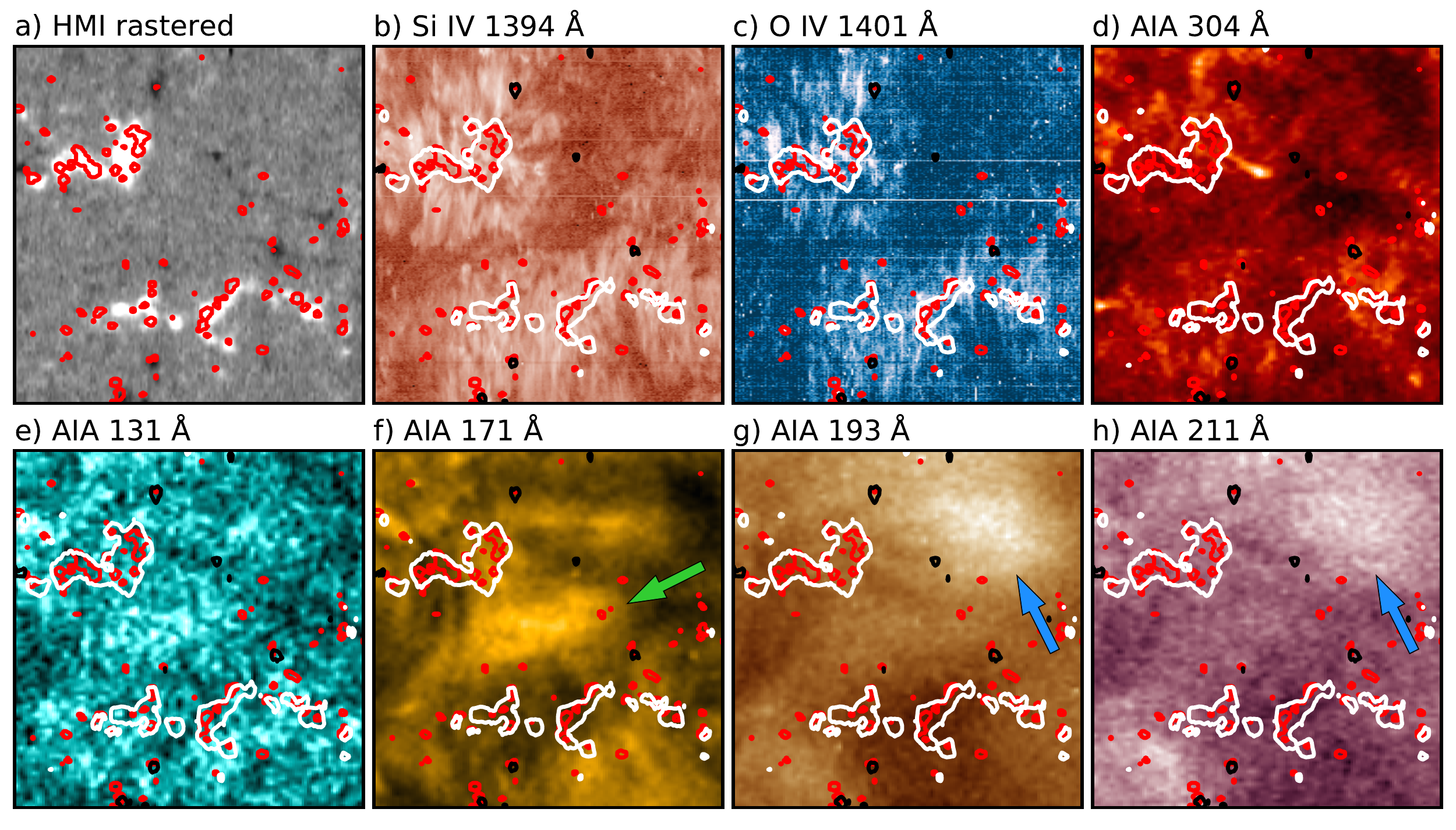}
\caption{Zoom into the diffuse region in different spectral bands. The individual panels represent a) the rastered line-of-sight magnetogram from HMI, as in \fig{F:data}c, b) the Si\,{\sc iv} 1394\,\AA\,emission line intensity from IRIS, as in \fig{F:data}a, c) the O\,{\sc iv} 1401\,\AA\,emission line intensity from IRIS, and d) -- h) observations from AIA in different channels indicated in each panel. Panels b) and c) are plotted in logarithmic scale. Like in \fig{F:data}, black and white contours outline the magnetic field at values of $-30$ G and $+30$ G, respectively. The red contours outline the brightenings in the Si\,{\sc i} Lyman continuum. The green arrow in panel f) points to the diffuse region we studied. The blue arrows in panels g) and h) point to a region with hotter plasma. See \sect{S:temp}.}
\label{F:temp}
\end{figure*}

\subsection{Magnetic field extrapolation\label{S:extrapolation}}

To relate the diffuse emission pattern we investigated to the large-scale magnetic field structure, we performed an extrapolation based on a potential field source surface (PFSS) model. To do this, we used the routines from the Python package pfsspy \citep{pfsspy}, which calculate the PFSS model for the given boundary conditions of the magnetic field. As boundary conditions, we used the synchronic map from HMI (the data product called hmi.mrdailysynframe\_polfil\_720s) mentioned at the end of \sect{S:obsdata}.

These data provide the magnetic field at the solar surface, with spatial pixels that are equally spaced in longitude and sine of latitude. We remapped these data to gain a spacing between the grid points of $0.5 ^{\circ}$ at the solar equator, which corresponds to about 6\,Mm. In the radial direction, we set a grid of 100 points equally spaced in logarithmic scale, and we set the radius of the source surface (where the magnetic field becomes radial) to be 2.5 $\mathrm{R_\odot}$. The resulting radial grid spacing is about 6.4\,Mm near the surface, similar to the horizontal spacing close to the solar equator. This is expected to be sufficient for our purpose.

After performing the extrapolation, we traced the magnetic field lines using routines from the same pfsspy Python package. The starting points for line-tracing, or "seeds", are located within and close to the IRIS field of view and always lie at a height of 0.015 $\mathrm{R_\odot}$ (about 10\,Mm) above the solar surface. The results of the extrapolation and field line tracing are discussed in \sect{S:largemagn}.

\section{Results\label{S:results}}

By combining the intensity images from the transition region and corona with the underlying magnetograms, we identified an unusual diffuse coronal feature. We use the temporal evolution and different wavelength bands and spectral lines to characterize this feature.

\subsection{Unusual diffuse coronal emission pattern\label{S:pattern}}

Emission from the quiet-Sun transition region originating from around 0.1\,MK is concentrated in the bright network. It forms bushel-type emission patterns that emerge from concentrations of the magnetic field, with elongated features pointing away from the magnetic field patches \cite[e.g.,][]{samanta2019}. This well-known solar feature is also seen in our data set in the Si\,{\sc iv} image in \fig{F:data}a.

Short cool loops are visible in Si\,{\sc iv} in the network lanes, in particular, in places with stronger patches of the magnetic field.  They extend for some 10 to 20\,Mm and are intermixed with short coronal loops seen in AIA 171\,\AA. In \fig{F:data} we highlight two such examples as squares 1 and 2 in the upper part of the field of view. As expected, these short loops are rooted in the footpoints of the opposite-polarity magnetic field (see \fig{F:data}a--c). They might be considered small versions of ephemeral regions or coronal bright points \cite[e.g.,][]{madjarska2019}.

However, the field of view also contains a region with an emission pattern that is quite different from the pattern described above (square 3 in \fig{F:data}). It shows two concentrations of positive magnetic field, which are highlighted with green arrows in \fig{F:data}c.  The emission patterns in Si\,{\sc iv} that emerge from these two positive polarities appear to be similar to the regions with spicular activity \cite[green arrows in \fig{F:data}a; e.g.,][]{samanta2019}. At the location between the two spicular regions in Si\,{\sc iv}, the AIA 171\,\AA\,images show a comparably bright diffuse-looking feature (green arrow in \fig{F:data}b). This is unusual, and it is the main focus of our study.

At first sight, the two spicular regions might be thought to be magnetically connected. Then the diffuse coronal emission, which extends to roughly 15\,Mm by 20\,Mm, would originate from the apex of this closed magnetic connection between the spicular footpoints. Because of the projection effects, the coronal emission would be visible between the bright spicular features in the transition region when observed from the top. The two spicular regions are located above magnetic field concentrations of the same (positive) magnetic polarity, however. Therefore, this scenario of closed magnetic loops is not applicable here.

\subsection{Temporal evolution of the diffuse region\label{S:evolution}}

The temporal evolution of this diffuse emission is another unusual characteristic. The diffuse emission persists for at least five hours, during which it evolves noticeably differently from the commonly observed magnetically closed (coronal) loops. Its brightness changes gradually, and certain parts of it become brighter or dimmer, but without obvious signs of (spatial) substructure. 

To investigate this evolution, we studied a sequence of AIA images in the 171\,\AA\,channel (\fig{F:evol}). The structure  of the diffuse region changes, but only gradually, and the region maintains its smooth appearance. This is best seen in the movie attached to \fig{F:evol}, but we also illustrate it with different snapshots of the diffuse region alone in \fig{F:evol}b. The variability of the average intensity in the diffuse region is shown in \fig{F:evol}c. It does not exceed about\,10\% over more than five hours. To emphasize that the changes are only gradual and the whole region remains quite diffuse-looking, we also created a space-time plot of the average intensity along a slit through the diffuse region. The space-time diagram is shown in \fig{F:evol}d, and the position of the slit is marked with AB in \fig{F:evol}b. It confirms that the evolution is gradual, without ridges or other features that would indicate dynamic evolution.

As exceptions of the rule, some locations show bright localized features. At time $t_1$ in \fig{F:evol}b, a bright feature is visible in the top left corner of the field of view. This appears to be associated with the magnetic field concentration at the same locations (see the white contours of the magnetic field in panel a). Another example is the bright compact feature at $t_4$ in \fig{F:evol}b below the middle of the field of view. This bright spot is also visible in the AIA 304\,\AA\,channel, which is imaging plasma at about 0.1\,MK. This indicates that this is a brightening in cooler transition region plasma that is also visible in the 171\,\AA\,channel. This channel images not only plasma just below 1 MK through emission from Fe\,{\sc ix} ions, but also transition region plasma at around 0.3\,MK from O\,{\sc v} ions (see \sect{S:temp}). Still, apart from exceptions like these, the region remains mainly diffuse over the course of five hours, as the movie attached to \fig{F:evol} shows.

Another diffuse-looking emission can also be found below square 1 in \fig{F:data}b. It is also located between spicular bushels with the same magnetic polarity at their footpoints (see \fig{F:data}a,c), just as the case we discussed in detail above. However, this second diffuse feature is closer to the bright point in square 1. Because this case is less isolated than the diffuse region discussed above, we did not focus on it as well here.

\subsection{Temperature of diffuse region\label{S:temp}}

The data from AIA alone are not sufficient to determine whether the diffuse region is a coronal or a transition region phenomenon.  We therefore added information from O\,{\sc iv} (1401\,\AA) acquired by IRIS to determine whether the diffuse emission forms around 1\,MK, or if it might originate from temperatures below 0.3\,MK.

The bright diffuse region is best visible in 171\,\AA\,(\fig{F:temp}d--h) of the different AIA channels. The feature is also recognizable, but less clear than in the 171\,\AA\,images, in the 131\,\AA\,filter, which is sensitive to emission from cooler plasma (see \tab{T.lines}). In the hotter channels, such as 193\,\AA\,and 211\,\AA, the region is fainter, but still visible\footnote{Another possibly hotter related structure is visible in 193\,\AA\,and 211\,\AA\, northwest of the diffuse region seen in 171\,\AA\,(blue arrow in \fig{F:temp}g--h). We did not analyze this structure here because we focus on the diffuse region in 171\,\AA\,images (green arrow in \fig{F:temp}f).}, while bright features in the 304\,\AA\,channel (from the transition region around 0.1\,MK) are mostly concentrated around the strong magnetic field patches in the network and generally avoid the area in which the diffuse region is located. 

Based on AIA images alone, the diffuse region is seen in channels that (predominantly) show plasma above 0.8 MK, but not at 0.1 MK. The temperature response of the 171\,\AA\,channel peaks around 0.85 MK (see \tab{T.lines}), which is mainly due to the Fe\,{\sc ix} lines. However, the transition region also contributes significantly at about 0.26\,MK \cite[e.g.,\,Fig.\,9 of][]{peter2022}, mostly from O\,{\sc v}.\footnote{\cite{peter2006} reported a formation temperature of 0.27\,MK for O\,{\sc v} lines in their magnetohydrodynamic model; see their Table 1.} Therefore, observations in 171\,\AA\,alone are not sufficient to determine whether the diffuse emission originates from a source region that is closer to transition region or coronal temperatures.

To narrow the temperature range of the diffuse region down, we used spectra of the O\,{\sc iv} (1401\,\AA) emission line, which forms at temperatures of about 0.15\,MK (see \tab{T.lines}) under the assumption of ionization equilibrium. In \fig{F:temp}c we show the observed intensity summed in a range of $\pm55$ km\,s$^{-1}$ around the rest wavelength of the line (20 spectral pixels). This image looks very similar to the Si\,{\sc iv} image. In particular, at the location of the diffuse region, both the O\,{\sc iv} and Si\,{\sc iv} images appear dark (\fig{F:temp}b,c). No prominent O\,{\sc iv} emission is visible. We therefore did not consider it likely that the diffuse emission in the 171\,\AA\,channel originates from O\,{\sc v}. Hence, we assumed that the observed diffuse plasma is closer to 1 MK. This is consistent with \cite{martinez-sykora2011}, who concluded that except for a few locations, the 171\,\AA\ channel is indeed dominated by emission at close to 1 MK.

\begin{figure*}
\centering
\includegraphics[width=\textwidth]{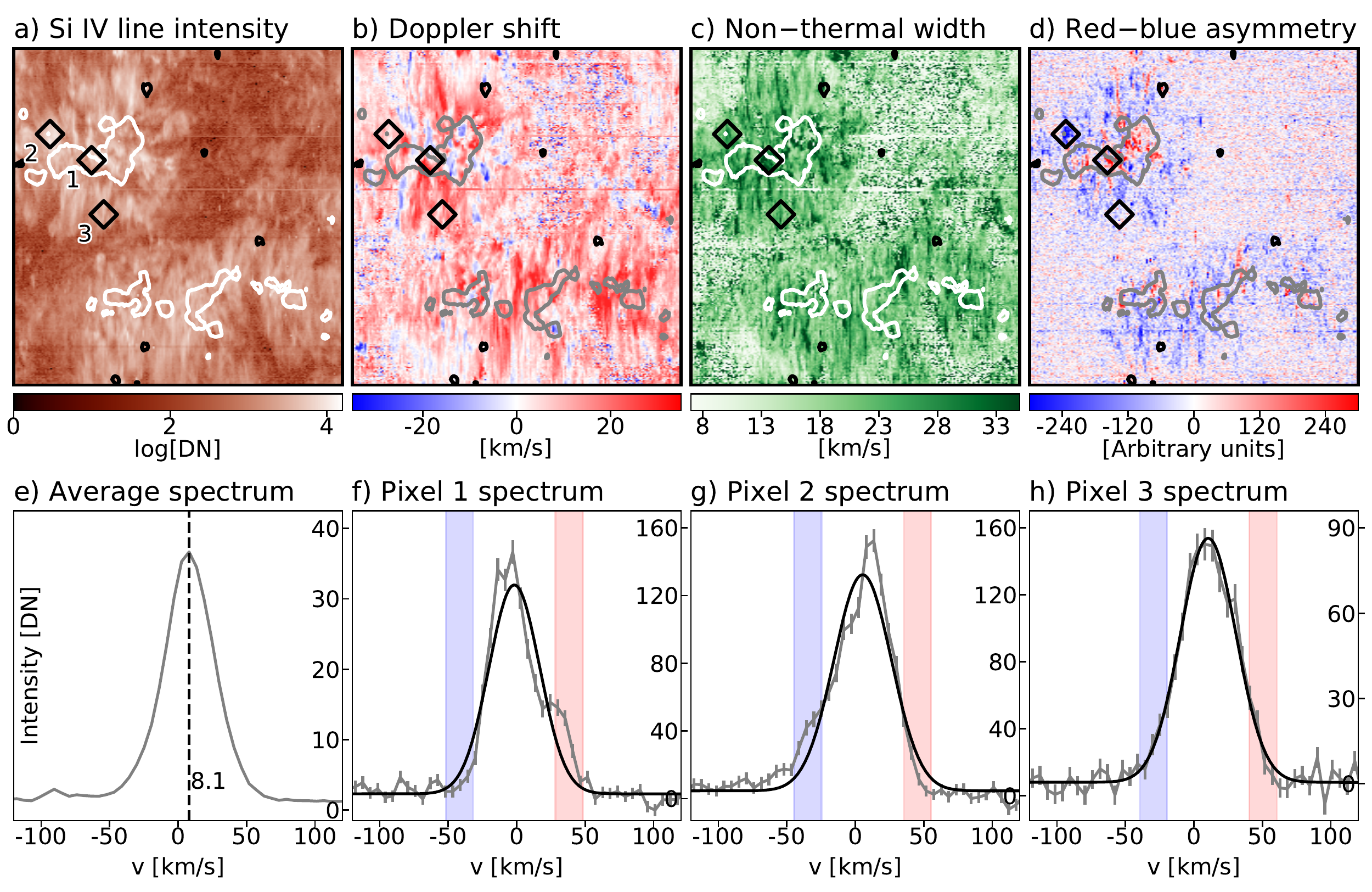}
\caption{Properties of the Si\,{\sc iv} 1394\,\AA\,line profiles in the vicinity of the diffuse region. The panels show a) the total line intensity, b) the Doppler shift, c) the nonthermal width,   d) the red-blue asymmetry, and e) the average profile of the whole area shown in the other panels. The average Doppler shift of 8.1\,km\,s$^{-1}$ is indicated. Panels f) -- h) show the profiles at three individual spatial pixels. The respective locations of these profiles are marked with diamonds in the top panels.  The gray bars in the profiles indicate the errors,  and black lines represent single Gaussians fitted to the profiles. The contours in the top panels have the same meaning as in \fig{F:data}. The wavelength is given in Doppler-shift units (positive is red) in the bottom panels. In panels f) -- h) the integration ranges used to calculate the red-blue asymmetry for the respective pixel are shaded. See \sect{S:spec}.}
\label{F:spec}
\end{figure*}

\begin{figure*}
\sidecaption
\includegraphics[width=120mm]{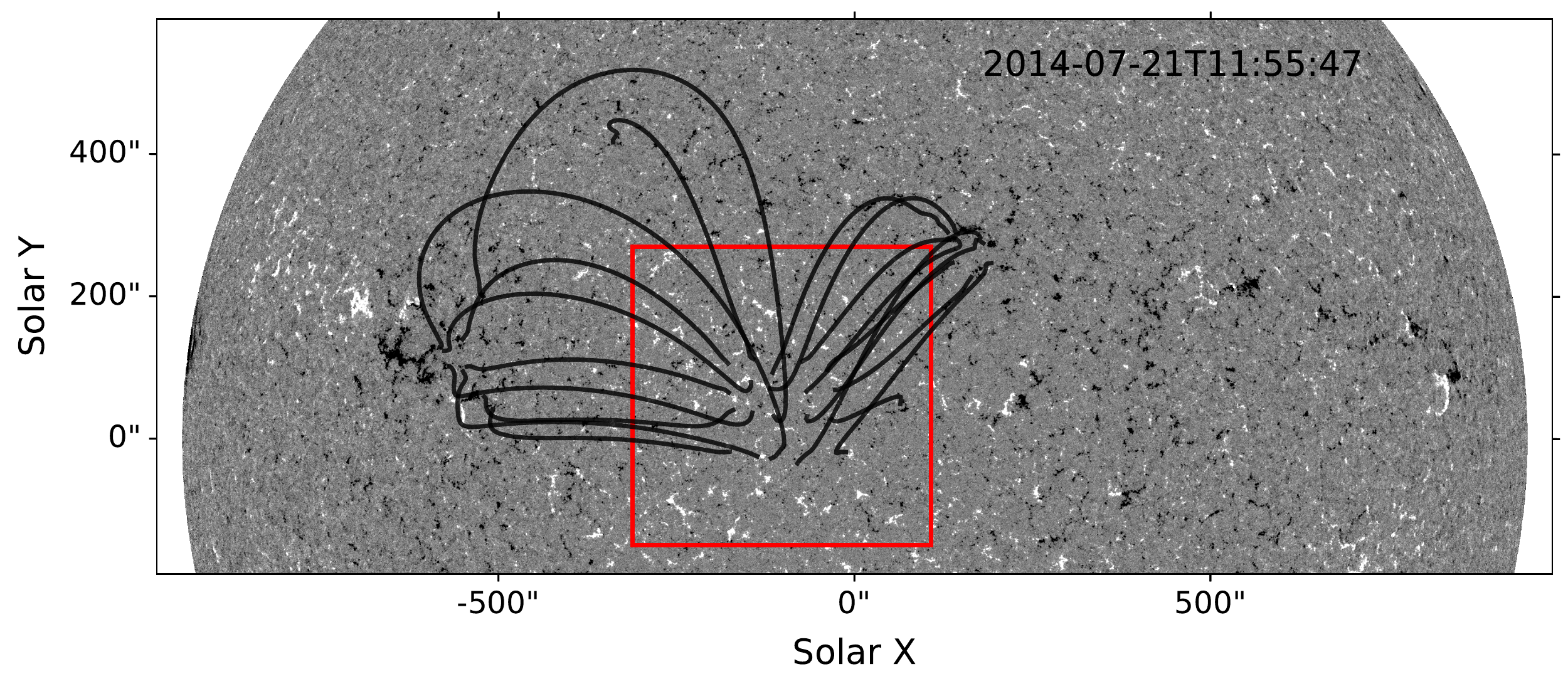}
\caption{Magnetic field configuration on a large scale related to the observed region of the diffuse emission. We show the long-range magnetic connections starting from the vicinity of the diffuse region calculated from a PFSS extrapolation. The background image is a line-of-sight magnetogram from HMI, in which the red box outlines the same area as the red box in \fig{F:cont}a. The diffuse region is located close to the center of the red box. See \sect{S:largemagn}.}
\label{F:largemagn}
\end{figure*}

\subsection{Si\,{\sc iv} line profiles at the spicular footpoints of the diffuse region\label{S:spec}}

We studied the Si\,{\sc iv} 1394\,\AA\,line profiles in the area around the diffuse region to investigate whether the two adjacent spicular regions are connected to the  diffuse coronal emission. To do this, we studied the line profile fits, and in particular, the line asymmetries, to determine the role of up- and downflows in and out of the diffuse region.

The results of the Gaussian line fits are summarized in \fig{F:spec}. Panels a) -- c) show the line intensity, Doppler shift, and nonthermal width. The average Doppler shift of the profiles in the field of view shown in \fig{F:spec}a is 8.1\,km\,s$^{-1}$ to the red (\fig{F:spec}e), which is largely consistent with the average value of the quiet-Sun transition region \cite[][]{peter1999}.

The Si\,{\sc iv} profiles are mostly single Gaussian, and, as usual for the lower transition region, they are predominantly redshifted (\fig{F:spec}b). However, many pixels still deviate from a single-Gaussian profile, with excursions to the red or the blue wing of the line. Examples of single pixels with and without obvious excursions are shown in panels f) -- h) of \fig{F:spec}.

To determine the spatial distribution of these excursions in the line wings, we calculated the red-blue asymmetry (RBA) of Si\,{\sc iv} at each spatial pixel as the difference between the integrals over spectral regions in the red and the blue wing \citep{mcintosh2009}. Prior to the integration, we linearly interpolated the spectrum to a grid with a spacing of about 0.9\,km\,s$^{-1}$. We chose this rescaling to create a spectral grid spacing that roughly matches the error of the line shift determined through a Gaussian fit. Then we calculated the line integrals in the range between [$+$30, $+$50]\,km\,s$^{-1}$ for the red and [$-$50, $-$30]\,km\,s$^{-1}$ for the blue wing, always with respect to the line center position deduced from the Gaussian fit. These spectral ranges were selected to roughly cover the line profile between one and two times the width of the line away from line center. To illustrate this, we indicate these ranges with the sample spectra from individual pixels in \fig{F:spec}, panels f) -- h). This choice shows the line asymmetry most clearly, even though the results do not critically depend on the exact choice of the spectral range. We show the RBA map obtained through this procedure in \fig{F:spec}d.

Based on the maps in \fig{F:spec}, stronger RBA values appear to be spatially well correlated with the larger nonthermal line widths. This is expected because excursions in line wings can be understood as separate components in the line profile, in addition to the main component. In a multicomponent line profile fit with a single Gaussian, the best fit is typically wider than the width of the individual components, hence the larger nonthermal widths at the locations of strong RBAs.

The features on the RBA map also resemble those of the intensity map. We predominantly see an excess emission in the blue wing in the magnetic network, in particular, at the edges between the network and the internetwork. This agrees with the concept from \cite{depontieu2009}, according to which, emission in the blue wing is expected when spicular bushels are observed that might feed material into the coronal structure. The excess in the red wing is mostly concentrated at the center of network elements, which may be indicative of fast downflows of cooling material from the coronal structures.

\subsection{Magnetic connectivity of the diffuse region\label{S:largemagn}}

The region we study is located in the quiet Sun. Its magnetic environment, which is based on the magnetic field distribution on the disk in \fig{F:cont}a, is puzzling, however. A larger area at the surface, outlined with the red box with a side length of 430\arcsec, is predominantly unipolar, with dominant positive-polarity magnetic flux concentrations (white).

Extended patches of unipolar regions like this are typically found at the base of coronal holes. In this case, however, AIA EUV images revealed no signature of coronal holes. Therefore, we expect that the magnetic field originating from our region of interest connects back to the surface. Because of the flux imbalance in this region, these magnetically closed connections must be long-ranging. The positive flux in this area originated from the decay of a large active region complex that was present there during multiple previous solar rotations. With the remnant magnetic flux of the decayed active regions, long-range magnetic connections there are expected.

To test this hypothesis quantitatively, we performed a magnetic field extrapolation based on a PFSS model (see \sect{S:extrapolation}). We show a sample of the traced field lines overplotted on a full-disk magnetogram from HMI in \fig{F:largemagn}. Closed field lines originate from the diffuse region with a footpoint distance of about 400\arcsec\,, which corresponds to about\,300\,Mm. Hence, the extrapolation confirms that the diffuse coronal region and the other locations about 300\,Mm away are connected.

\section{Discussion\label{S:discuss}}

\begin{figure}
\centering\includegraphics[width=70mm]{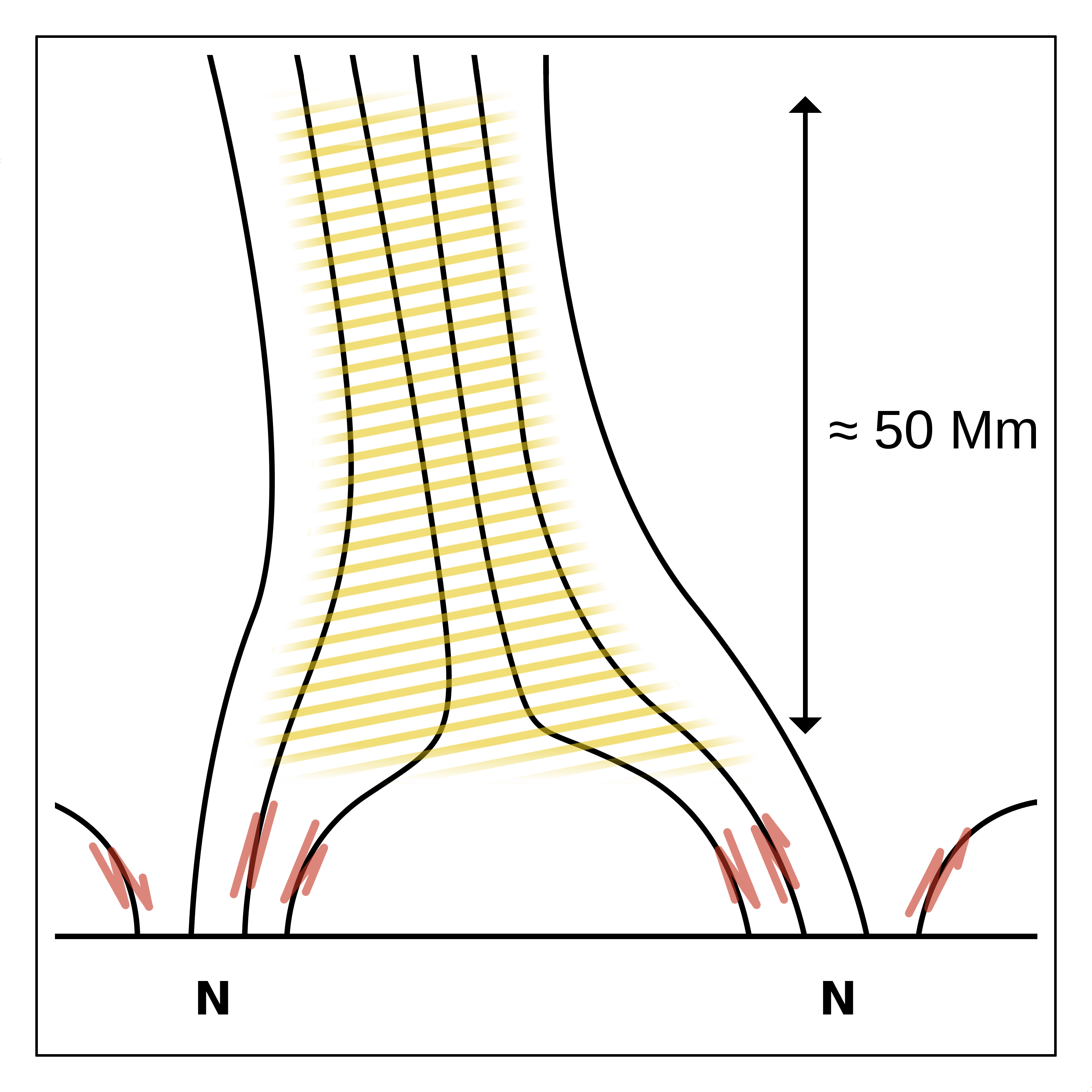}
\caption{Relation of the diffuse emission to the magnetic field configuration. We illustrate the vertical cross section of the observed system along the direction that connects the two positive magnetic field concentrations from \fig{F:temp}a. The magnetic field patches are marked with N for the northern polarity, and the spicular emission they host is shown in orange. The magnetic field lines form a canopy below the source region of the diffuse emission. This is shown in yellow. See \sect{S:magn}.}
\label{F:magn}
\end{figure}

\subsection{Magnetic structure of diffuse coronal emission\label{S:magn}}

Its placement between two spicular bushels with magnetic footpoints of the same polarity is one of the puzzling characteristics of the observed diffuse emission. In the following, we discuss one possible scenario in which the diffuse emission is related to the configuration of the magnetic field.

It is reasonable to assume that the footpoints of the diffuse emission are indeed the two neighboring spicular bushels, both rising from concentrations of the similar-polarity magnetic field (\fig{F:temp}a). Hence, we expect that the local magnetic configuration looks similar to the cartoon picture in \fig{F:magn}. This is essentially inspired by the scenario outlined by \cite{depontieu2009}, who suggested that coronal loops are fed through spicules. The cartoon shows a vertical cross section through the diffuse coronal region with its magnetic footpoints. The latter are marked N, for northern (positive) magnetic polarity. The spicular features in the transition region are indicated in orange. Because both patches are of the same magnetic polarity, they form a magnetic canopy above and between them that extends farther into the corona. The plasma that radiates apparently diffuse coronal emission (yellow) is then located within this canopy between the spicular bushels. It remains unclear whether the coronal plasma on these field lines is loaded through the spicules or through some other process. 

The diffuse emission we observe is present for many hours, which is longer than the typical coronal cooling time \cite[shorter one hour for 1\,MK plasma; e.g.,][]{aschwanden2005}. This should be enough time for this structure to reach some near-equilibrium state and be close to a barometric stratification. It should extend up vertically to at least one pressure scale height, which corresponds to about 50\,Mm for plasma of about\,1\,MK. The diffuse feature is observed as a nearly compact structure near disk center, however, located between network patches of the same magnetic polarity (see \fig{F:temp}f; green arrow). For it to be an extended feature that appears compact to us, it therefore has to be embedded in a magnetic field structure that is close to vertical up to a height of about 50\,Mm (see the cartoon in \fig{F:magn}).

In \sect{S:largemagn} we reported that the diffuse region is located within an area with long-ranging magnetic connection. Therefore, the approximately\,50\,Mm tall vertical magnetic structure could be a part of one of these long connections (loops). When we assume a semicircular field line that closes at 300\,Mm away from the diffuse region, a structure at one end of this semicircle spanning a height of 50\,Mm would be almost vertical. In the projection along the vertical, simple geometry tells us that at a height of 50\,Mm, the field line would have moved horizontally away from the footpoint by only 8.5\,Mm. We would therefore expect a horizontal extent of the diffuse region to be approximately 10\,Mm. This is roughly consistent with the observed extent of the diffuse region (see \fig{F:data}b and the scale in panel a). Therefore, the diffuse emission we observe might originate from plasma in the stem of an long-ranging loop.


One key to understanding the diffuse look of the region might be found in its height. If the scenario put forward here (and in the cartoon in \fig{F:magn}) is indeed applicable, then the long line of sight would quite naturally result in averaging individual features within the volume of the structure. To further investigate this, observations from two vantage points close to quadrature would be highly desirable.

\subsection{Supply of hot gas to the diffuse corona above the quiet Sun}

We interpreted the temperature and the magnetic structure of the diffuse emission, but the question of its formation mechanism still remains. We speculate about possible processes here that have to be evaluated and confirmed or refuted in future studies.

It seems natural to search for explanations related to the magnetic footpoints. The spectra of the Si\,{\sc iv} line from IRIS show excess emission in the blue wing above the magnetic field concentrations, in particular, south of the diffuse coronal patch (see \figs{F:temp}f and \ref{F:spec}d). This suggests upflows that might be filling the region of the diffuse emission with hot plasma. These upflows might be driven by the heating processes in the lower layers, as in the case of type II spicules, if they really reached coronal temperatures \cite[][]{depontieu2017}. Otherwise, they might also be driven by coronal heating events. In this traditional scenario, heat conduction causes chromospheric evaporation at the footpoints \cite[][]{patsourakos2006}. To distinguish between these two processes, that is, spicule injection versus\,evaporation, we would need to have access to the temporal evolution in Si\,{\sc iv} as well, and also to spectral information from the hotter coronal plasma at about one million Kelvin, neither of which is available here. 

Our findings here indicate upflows in the footpoint area that might be related to spicules. These upflows may not be related to the diffuse emission, however. The apparent relation could just be coincidence. Furthermore, the region we observed, although it lies in the quiet Sun, is reminiscent of coronal holes because of its predominantly unipolar magnetic flux at the surface. Plumes that appear hazy have been observed in coronal holes before. They show highly dynamical activity at their footpoints \cite[e.g.,][]{raouafi2014}. This activity, in the form of jetlets and plume transient bright points, is thought to be caused by cancellation of the minority polarity with the dominant unipolar flux. It could be important for the sustainability of the plumes for time periods of hours to days. Following this analogy, the activity at the footpoints of the diffuse region we observed might be important for its formation and might support it for several hours.

The diffuse emission is also part of a long-ranging magnetic connection. This opens the possibility that the plasma producing the diffuse emission is being supplied from flows originating from the farther footpoint (the other side of the long-range connection) instead of being supplied by the upflows from the closer footpoint. Downflows in the transition region above sunspots have been observed many times before; these are the footpoints of the long coronal loops \cite[][]{chitta2016, chen2022}. The downflows are detected as redshifted components in transition region emission lines, often with supersonic speeds. To support these downflows, material must be supplied from the other footpoint, possibly via a siphon flow. To verify whether the diffuse emission patch is caused by supersonic downflows sustained by upflows from the conjugate footpoint, we investigated the average profiles of Si\,{\sc iv} (1394\,\AA) and O\,{\sc iv} (1401\,\AA). We did not find any significant redshifts, however, which suggests that there are no indications of downflows, at least at transition region temperatures, at the location of the diffuse region. EUV spectroscopic observations covering higher temperatures around 1\,MK may help to provide better information about the nature of the plasma flows in this type of region.

With the observations presented here, the relevant processes that feed plasma into the diffuse region remain an open question. Access to data with a temporal evolution of the transition region and the about 1 MK hot coronal plasma, and spectral data in both domains, would provide better constraints on the formation mechanism of similar regions.

\section{Conclusions\label{S:concl}}

We reported an observation of the diffuse emission on supergranular scales, seen on-disk in AIA images of close to 1 MK plasma (\fig{F:temp}). It persisted for many hours and was not reminiscent of small loops, which are typical for the upper transition region and lower corona (\fig{F:evol}). It remained diffuse and looked cloudy, even though the underlying atmosphere shows various small features. Unlike a typical small loop, it was not located directly above the magnetic network, but in the internetwork area between two concentrations of positive magnetic field (\fig{F:data}).

Based on our findings, we suggest that we observed diffuse emission that originated from long-ranging loops with highly structured spicular footpoints at their base. Essentially, we observed the (mostly) vertical part of long loops over one pressure scale height above the footpoints. This means that the line-of-sight integration might cause the region to appear diffuse. The spicule-type features at the base of the diffuse coronal emission suggest spicules as one possible mechanism that might feed the diffuse region with plasma, but evaporation might also provide the material for the diffuse coronal patch. Mass fed into the diffuse region by a siphon flow from the farther footpoint seems less likely. Within the scope of this paper, the true formation mechanism of the observed diffuse emission remains elusive, however.

This setup of an area with one dominant magnetic polarity that is engaged in long-range connections is intriguing. The long-term evolution of the magnetic field (over more than one solar rotation) shows that in our case, this dominance of one polarity in a large region is a remnant from (part of) a complex active region. These extended patches of unipolar regions commonly reflect the long-term decay of active regions, which means that the associated long-ranging magnetic connections between the decaying active regions and the quiet Sun might contribute substantially to the overall coronal emission.

Diffuse features on supergranular scales, similar to the feature studied here, can be found regularly in the solar corona. In the future, we plan to observe them in quadrature with SDO and the remote-sensing instruments on board the Solar Orbiter mission \citep{SOLO}. The simultaneous view on the disk and at the limb from the two platforms will provide us with more information about their shape and vertical extension.


\begin{acknowledgements}
We greatly appreciate the constructive comments from the referee.
This work was supported by the International Max-Planck Research School (IMPRS) on Physical Processes in the Solar System and Beyond.
N.M. acknowledges a PhD fellowship of the International Max Planck Research School on Physical Processes in the Solar System and Beyond (IMPRS). L.P.C. gratefully acknowledges funding by the European Union (ERC, ORIGIN, 101039844). Views and opinions expressed are however those of the author(s) only and do not necessarily reflect those of the European Union or the European Research Council. Neither the European Union nor the granting authority can be held responsible for them.
\end{acknowledgements}

\end{document}